\documentclass[aps,prl,superscriptaddress,amsmath,amssymb,twocolumn,amsfonts,floatfix, longbibliography,10pt]{revtex4-2}

\usepackage{chemformula,siunitx}
\usepackage{lastpage}
\usepackage[protrusion=true, expansion=true]{microtype}
\usepackage{wrapfig}
\usepackage{multirow}
\usepackage{array}
\usepackage{ragged2e}
\justifying

\usepackage{graphicx}
\usepackage{epstopdf}
\usepackage[T1]{fontenc}
\usepackage{amsbsy}
\usepackage[T1]{fontenc}
\usepackage{amsmath}
\usepackage{amssymb}
\usepackage{bbm}
\usepackage{braket}
\usepackage{xcolor}
\allowdisplaybreaks
\usepackage{graphicx}

\usepackage[normalem]{ulem}

\newcommand{\tableref}[1]{Table~\ref{#1}}

\renewcommand{\approx}{\simeq}

\renewcommand{\vec}[1]{\boldsymbol{#1}}

\usepackage[colorlinks=true]{hyperref}
\hypersetup{
    unicode=false,          
    pdftoolbar=true,        
    pdfmenubar=true,        
    pdffitwindow=false,     
    pdfstartview={FitH},    
    pdftitle={PtPb3Bi},    
    pdfauthor={Shashank Srivastava},      
    pdfsubject={},   
    pdfcreator={},   
    pdfproducer={}, 
    pdfkeywords={} {} {}, 
    pdfnewwindow=true,      
    colorlinks=true,       
    linkcolor=blue, 
    citecolor=blue,        
    filecolor=magenta,      
    urlcolor=blue          
}
\usepackage{orcidlink}
\begin{document}

\title{Discovery of Quasi-One-Dimensional Superconductivity in \ch{PtPb3Bi}}

\author{{Shashank Srivastava}\,\orcidlink{0009-0009-5065-4516}}
\affiliation{Department of Physics, Indian Institute of Science Education and Research Bhopal, Bhopal, 462066, India}

\author{{Yash Vardhan}\,\orcidlink{0009-0003-7290-5300}}
\affiliation{Department of Condensed Matter Physics and Materials Science, Tata Institute of Fundamental Research, Colaba, Mumbai 400005, India}

\author{Anshu Kataria}
\affiliation{Department of Physics, Indian Institute of Science Education and Research Bhopal, Bhopal, 462066, India}

 \author{Pradyumna Bawankule}
\affiliation{Department of Physics, Indian Institute of Science Education and Research Bhopal, Bhopal, 462066, India}

\author{Poulami Manna}
\affiliation{Department of Physics, Indian Institute of Science Education and Research Bhopal, Bhopal, 462066, India}

\author{Prabin Kumar Naik}
\affiliation{Department of Physics, Indian Institute of Science Education and Research Bhopal, Bhopal, 462066, India}

\author{{Rahul Verma}\,\orcidlink{0009-0008-7902-6866}}
\affiliation{Department of Condensed Matter Physics and Materials Science, Tata Institute of Fundamental Research, Colaba, Mumbai 400005, India}

\author{Rhea Stewart}
\affiliation{ISIS Facility, STFC Rutherford Appleton Laboratory, Oxfordshire, OX11 0QX, United Kingdom}

\author{James S. Lord}
\affiliation{ISIS Facility, STFC Rutherford Appleton Laboratory, Oxfordshire, OX11 0QX, United Kingdom}

\author{Adrian D. Hillier}
\affiliation{ISIS Facility, STFC Rutherford Appleton Laboratory, Oxfordshire, OX11 0QX, United Kingdom}

\author{Mathias~S.~Scheurer}
\affiliation{Institute for Theoretical Physics III, University of Stuttgart, 70550 Stuttgart, Germany}

\author{D. T. Adroja}
\affiliation{ISIS Facility, STFC Rutherford Appleton Laboratory, Oxfordshire, OX11 0QX, United Kingdom}
\affiliation{Highly Correlated Matter Research Group, Physics Department, University of Johannesburg, Auckland Park 2006, South Africa}

\author{{Bahadur Singh }\,\orcidlink{0000-0002-2013-1126}}
\email[]{bahadur.singh@tifr.res.in}
\affiliation{Department of Condensed Matter Physics and Materials Science, Tata Institute of Fundamental Research, Colaba, Mumbai 400005, India}

\author{{Ravi Prakash Singh}\,\orcidlink{0000-0003-2548-231X}}
\email[]{rpsingh@iiserb.ac.in}
\affiliation{Department of Physics, Indian Institute of Science Education and Research Bhopal, Bhopal, 462066, India}

\begin{abstract}
Quasi-one-dimensional (quasi-1D) materials provide a compelling platform where reduced dimensionality stabilizes intertwined topological and superconducting phases. Here we report superconductivity in a new Bi-based quasi-1D compound, \ch{PtPb3Bi}, which hosts a nontrivial electronic structure. It exhibits type-II superconductivity below 3.01(1) \si{K}. Heat capacity and transverse-field muon spin rotation/relaxation ($\mu$SR) measurements demonstrate a fully gapped isotropic $s$-wave state with moderate electron-phonon coupling, while zero-field $\mu$SR confirms the preservation of time-reversal symmetry (TRS). Transport measurements reveal low carrier mobility with diffusive normal-state transport. Electronic structure calculations show strong dispersion along the quasi-1D direction and relatively flatter bands in the transverse plane, giving rise to pronounced Fermi surface nesting in the $k_x$–$k_y$ plane. Consistent with this, the compound undergoes a charge-density-wave transition at 280(1) \si{K}. The flow of Wannier charge centers, together with surface-state dispersion, establishes nontrivial band topology. These results identify \ch{PtPb3Bi} as a new quasi-1D superconductor with nontrivial electronic structure and a promising candidate for topological superconductivity.
\end{abstract}

\maketitle

\noindent \textit{Introduction:}
Low-dimensional materials have attracted widespread interest owing to emergent quantum phenomena uniquely enabled by reduced dimensionality~\cite{Zhang01062007,Li0.9Mo6O17}. Quasi-one-dimensional (quasi-1D) materials represent an extreme limit, where reduced crystalline dimensionality gives rise to highly anisotropic electronic structures and a strong propensity for exotic quantum states~\cite{nontrivial2020tanite5,topological2022Bi4X4}. In such systems, the interplay of disorder and electron correlations is enhanced, providing a fertile platform to explore localization and interaction effects that are difficult to access in higher dimensions~\cite{petrovic2016disorder,manna2026evidence,nonsymm2021TaSe3}. Despite this, superconductivity in quasi-1D materials remains relatively rare at ambient pressure, as the strong tendency toward charge density wave (CDW) formation -- driven by Fermi surface nesting and electron-phonon coupling -- gaps large portions of the Fermi surface and suppresses superconductivity. Moreover, reduced dimensionality enhances thermal and quantum fluctuations, further destabilizing long-range superconducting phase coherence~\cite{pressurekmn6bi5,pressurequasi1D,pressureta2pdse6}.

Recent advances have highlighted the importance of nontrivial electronic structures in superconducting materials, where the coexistence of superconductivity and band topology provides a potential route to unconventional superconducting states~\cite{Sato,taptsi,nayak2008topo,RhGe_SM}. While some quasi-1D superconductors have been proposed to host nontrivial topology, most of the topological superconductor candidates identified so far are based on 2D or 3D materials~\cite{nonsymm2021TaSe3,nontrivial2020tanite5,V2Ga5,topological2022Bi4X4}. This highlights the need for systematic exploration of quasi-1D materials in which superconductivity coexists with nontrivial electronic structure and for understanding how such topology influences the superconducting pairing mechanism.

\begin{figure*}[t]
\includegraphics[width=0.9\textwidth]{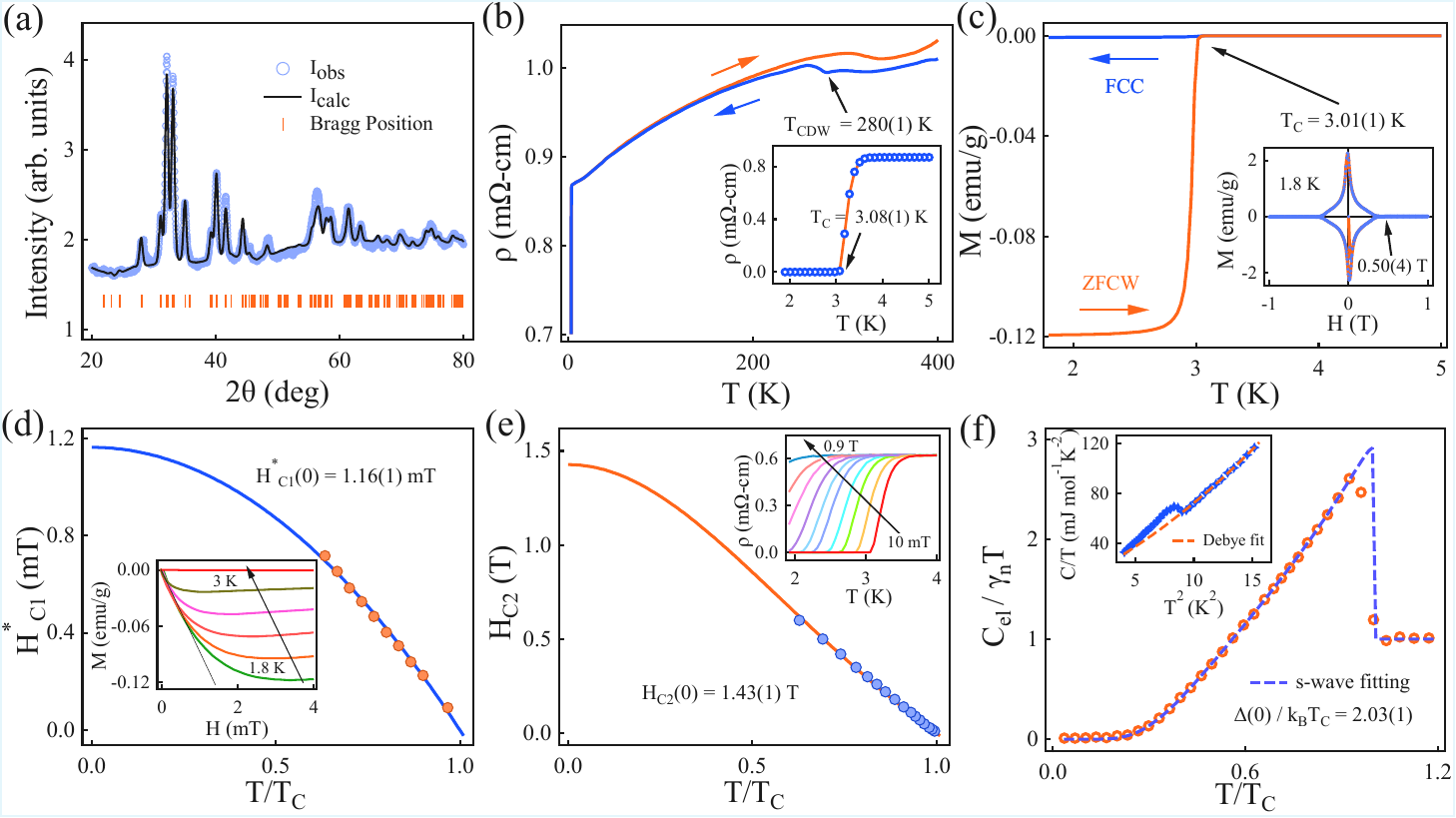}
\caption {\label{Fig:mag} \textbf{Structural and superconducting characterization of \ch{PtPb3Bi}:}
(a) Rietveld refined room temperature powder XRD.
(b) Resistivity $\rho(T)$ versus temperature at zero magnetic field. A CDW transition at 280(1) \si{K} and a superconducting transition at 3.08(1) \si{K} (left inset) is revealed. 
(c) Variation of magnetic moment with temperature in ZFCW and FCC mode measured under an applied magnetic field of 1 \si{mT}. Inset shows the five-quadrant magnetic field dependence of the magnetic moment.
(d) Lower critical field and (e) upper critical field variation with reduced temperature ($T/T_C$) fitted using GL equation to estimate $H^*_{C1}$(0) and $H_{C2}$(0), respectively. Insets in (d) and (e) show magnetic moment versus magnetic field at different temperatures from 1.8 \si{K} to 3 \si{K}, and temperature-dependent resistivity measured under fields from 10 \si{mT} to 0.9 \si{T}.
(f) Normalized electronic specific heat versus $T/T_C$ fitted using the isotropic $s$-wave model. Inset: Variation of $C/T$ with $T^2$ at zero magnetic field is well described using the Debye relation.
}
\end{figure*}

In this context, we investigate a new Bi-based quasi-1D superconductor, \ch{PtPb3Bi}, as a platform where superconductivity coexists with nontrivial electronic structure. 
Bi-based superconductors are known to constitute promising platforms for supporting topological surface states and facilitating the emergence of Majorana zero modes~\cite{sharma2024gammabipd,sakano2015Bi2Pd,dimitri2018alphaBi2Pd,Bi2PdPt}. The choice of \ch{PtPb3Bi} is motivated by related Pt-Pb and Pt-Sn compounds, such as \ch{PtPb4} and \ch{PtSn4}, which exhibit topological semimetallic behavior~\cite{xu2021superconductivity,Li2019PtSn4}. Moreover, its quasi-1D crystal structure and symmetry (D$_{4h}$) provide a setting where unconventional superconducting states can, in principle, emerge~\cite{Lu5Rh6Sn18,shang2020remo,sakano2015Bi2Pd}. The point group D$_{4h}$ has two 2D irreducible representations (IR), $E_g$ and $E_u$, and hence allows for TRS-breaking superconductivity at a single phase transition \cite{RevModPhys.63.239}.

Here, using in-depth magnetization, electrical resistivity, heat capacity, and $\mu$SR measurements, we establish bulk superconductivity in \ch{PtPb3Bi} with a fully gapped isotropic $s$-wave state, moderate electron-phonon coupling, and preserved time-reversal symmetry. We further identify a CDW transition at 280(1) \si{K} and highly disordered nature consistent with quasi-1D nature. Electronic structure calculations reveal pronounced dispersion along the quasi-1D direction ($z$-axis), relatively flatter dispersion along the in-plane direction with Fermi-surface nesting, and nontrivial surface states, establishing \ch{PtPb3Bi} as a robust platform to investigate the interplay of superconductivity, reduced dimensionality, and nontrivial band topology.

\begin{figure*}[t]
\centering
\includegraphics[width=0.90\textwidth]{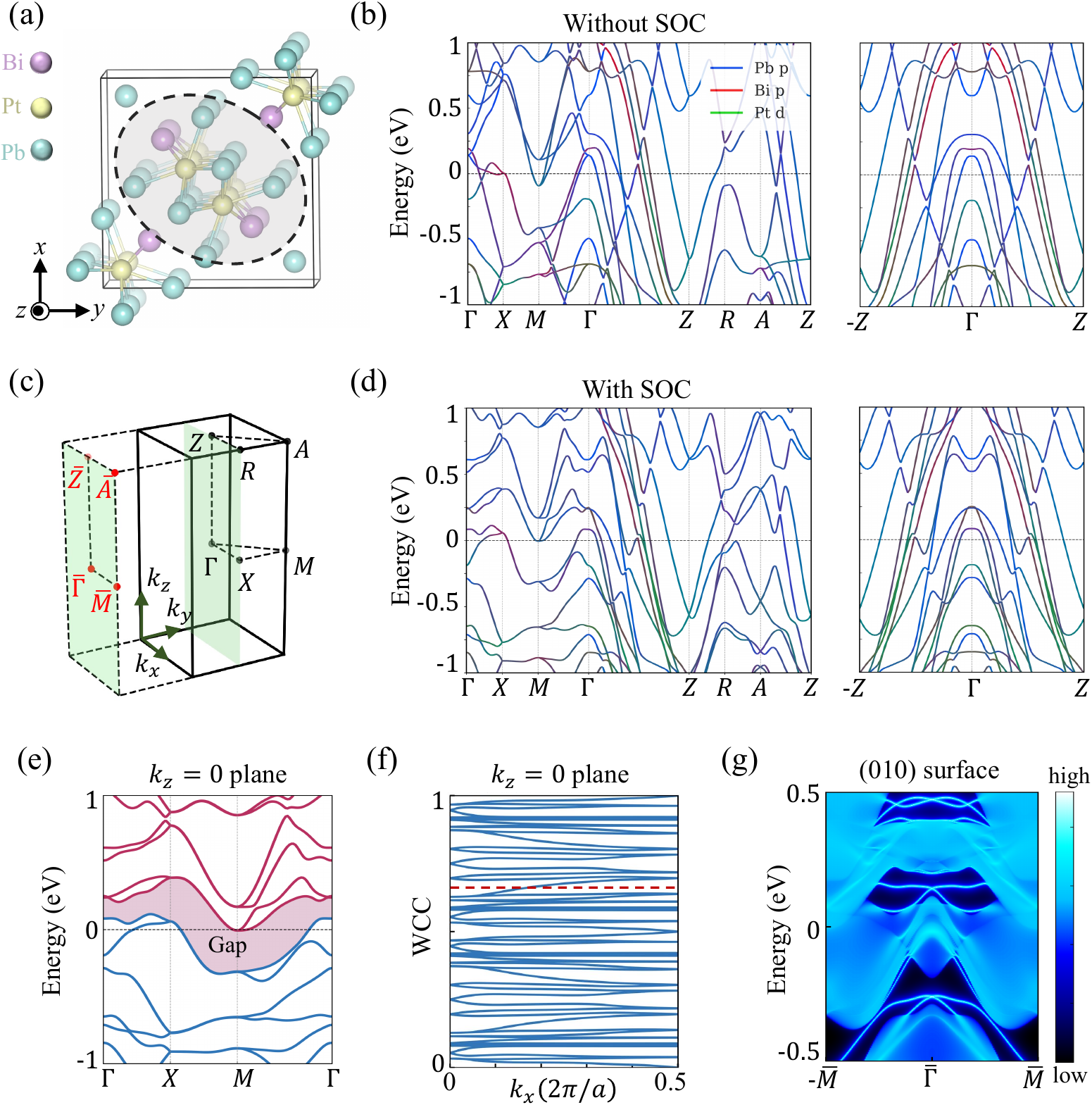}
\caption{\label{Fig:DFT} \textbf{Electronic band structure of \ch{PtPb3Bi}:}
(a) Crystal structure of \ch{PtPb3Bi}, with quasi-1D chains along the $z$ axis.
(b) Orbital-resolved band structure without SOC. Left: dispersion along high-symmetry directions in the full Brillouin zone (BZ); right: band dispersion along the quasi-1D $k_z$ direction, where highly dispersive bands cross the Fermi level.
(c) Bulk BZ with high-symmetry points; the (010) surface-projected BZ is also shown.
(d) Same as (b), but with SOC, where gaps open at multiple band crossings and band dispersion changes.
(e) Bulk band structure on the $k_z = 0$ plane, with a finite continuous band gap highlighted.
(f) Evolution of Wannier charge centers (WCCs) at the $k_z = 0$ plane, indicating a nontrivial $\mathbb{Z}_2$ topological invariant.
(g) (010) surface band structure. The nontrivial surface states appear in the bulk gap, with a Dirac point at $\overline{\Gamma}$.
}
\end{figure*}

\noindent \textit{Sample characterization and superconducting properties:}
Polycrystalline sample \ch{PtPb3Bi} was synthesized using the solid-state reaction technique. Rietveld refinement of the powder XRD confirms that \ch{PtPb3Bi} crystallized in a single phase (Fig. \ref{Fig:mag}(a)) in the tetragonal space group \textit{P4$_2$/mnm} (No.~136). The crystal possesses inversion symmetry ($\mathcal{I}$), a two-fold rotation axis along the $z$-direction ($\mathcal{C}_{2z}$), and a mirror plane perpendicular to the $z$-axis ($\mathcal{M}_z$). In addition, it hosts several nonsymmorphic symmetries, including $\{\mathcal{C}_{2x}|\tfrac{1}{2},\tfrac{1}{2},\tfrac{1}{2}\}$, $\{\mathcal{C}_{2y}|\tfrac{1}{2},\tfrac{1}{2},\tfrac{1}{2}\}$, and $\{\mathcal{C}_{4z}|\tfrac{1}{2},\tfrac{1}{2},\tfrac{1}{2}\}$. The crystal structure consists of linear chains of Pt and Bi atoms along the $z$ direction, interconnected by Pb atoms (see Fig.~\ref{Fig:DFT}(a)). This arrangement forms a quasi-one-dimensional (quasi-1D) network of Pt--Bi--Pb chains along the $z$ direction.

DC magnetization, zero-field specific heat, and AC resistivity data confirm superconductivity in \ch{PtPb3Bi}. Resistivity sharply drops to zero at 3.08(1) \si{K} (Fig. \ref{Fig:mag}(b), inset). The resistivity data exhibit a CDW transition at 280(1) \si{K} (Fig. \ref{Fig:mag}(b)), which probably emerged from the quasi-1D structure and Fermi surface nesting (see below). A strong diamagnetic transition is observed in magnetization data at 3.01(1) \si{K} as shown in Figure~\ref{Fig:mag}(c). The strong pinning in field-cooled cooling (FCC) data, along with the M-H loop at 1.8 \si{K} (Fig. \ref{Fig:mag}(c), inset), suggests type-II superconductivity. Figures~\ref{Fig:mag}(d) and (e) show the lower and upper critical fields ($H_{C1}=2.09(4)$ \si{mT} and $H_{C2}=1.43(1)$ \si{T}), determined from the Ginzburg-Landau (GL) fitting of the temperature dependent magnetization versus magnetic field data, and magnetic field dependent resistivity versus temperature data, respectively (see SM for details \cite{SM}). Using GL theory, the superconducting length parameters were estimated to determine the GL parameter $\kappa_{\text{GL}}$. The coherence length $\xi_{\text{GL}}=15.8(1)$ \si{nm} and penetration depth $\lambda_{\text{GL}}=534.5(5)$ \si{nm}, gives $\kappa_{\text{GL}}=33.8(4)$ confirming type-II superconductivity in \ch{PtPb3Bi}. Low value of the Maki parameter $\alpha_{m}=0.23(2)$ shows that the upper critical field is dominated by the orbital limiting effect over the Pauli limiting effect. Bulk superconductivity was supported by the superconducting jump in the zero-field specific heat at 2.96(4) \si{K}. Normal region of the low-temperature specific heat data was fitted with the Debye-Sommerfeld relation (see the inset of Fig. \ref{Fig:mag}(f)). Electronic specific heat data best fit with the $s$-wave model (Fig. \ref{Fig:mag}(f)), yielding a superconducting energy gap of 2.03(1). Calculated value of $\frac{\Delta C_{\text{el}}}{\gamma_{n}T_C}$ and electron-phonon coupling constant $\lambda_{\text{e-ph}}$ are 1.64(2) and 0.71(3), respectively, indicating moderate strength of electron-phonon interaction. Hall measurements were carried out to determine the carrier mobility of the sample. Calculated mobility is 0.35(1) \si{cm^2V^{-1}s^{-1}}, indicating very low charge carrier transport in the material. Details of these measurements and the calculated superconducting parameters are discussed in SM \cite{SM}.

\noindent \textit{Electronic Structure:}
Bulk Brillouin zone with high-symmetry points and its (010) surface projection are shown in Fig.~\ref{Fig:DFT}(c). The orbital-resolved bulk band structure of \ch{PtPb3Bi} is presented in Figs.~\ref{Fig:DFT}(b) and (d). Without SOC, multiple bands cross the Fermi level. The bands show high dispersion along $-Z$-$\Gamma$-$Z$, while they remain relatively flat along the in-plane $k_x$-$k_y$ directions, consistent with the quasi-1D structure along $z$. States near the Fermi level derive primarily from Pb $p$, Bi $p$, and Pt $d$ orbitals, which cross the Fermi level and indicate metallic behavior. Bands along $k_z$ show a parabolic dispersion spanning approximately $-1$~\si{eV} to $2$~\si{eV}, whereas flatter bands in the $k_x$–$k_y$ plane indicate lower in-plane mobility. This result supports the experimentally observed low carrier mobility in \ch{PtPb3Bi}. Including SOC opens gaps at several crossings near the Fermi level and modifies the band dispersion (Fig.~\ref{Fig:DFT}(d)).

\ch{PtPb3Bi} exhibits a continuous gap across the $k_z = 0$ plane, as highlighted in Fig.~\ref{Fig:DFT}(e). The presence of this finite gap allows evaluation of the topological invariant in a manner similar to insulators. The Wannier charge center (WCC) evolution shows an odd number of crossings of a reference line, which indicates a nontrivial $\mathbb{Z}_2 = 1$ state (Fig.~\ref{Fig:DFT}(f)). Moreover, surface calculations along the (010) plane confirm this result. Nontrivial surface states appear within the bulk band gap, as shown along $-\bar{M}$-$\bar{\Gamma}$-$\bar{M}$ in Fig.~\ref{Fig:DFT}(g), consistent with the nontrivial bulk topology.

\begin{figure}[t]
\centering
\includegraphics[width=\columnwidth]{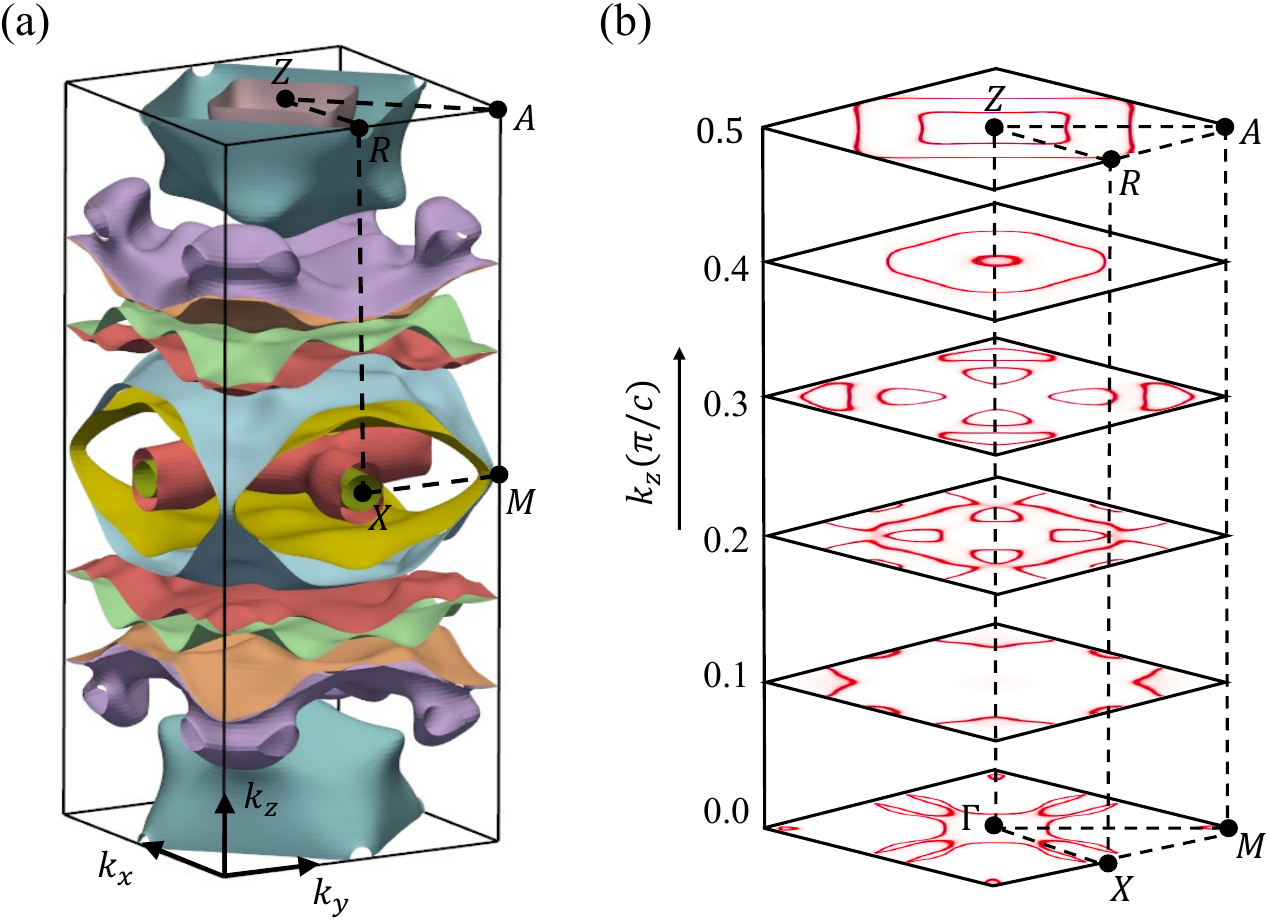}
\caption{\textbf{Fermi surface and Fermi band contour:} (a) Fermi surface of \ch{PtPb3Bi} in the bulk Brillouin zone. (b) Fermi band contour along $k_z$, showing evolution of Fermi contours with $k_z$. Nearly planar contours appear at $k_z=\pi/c$ plane.}
\label{Fig:FS}
\end{figure}

Figure~\ref{Fig:FS}(a) shows the calculated Fermi surface, which consists of multiple pockets and sheets. Fermi band contours in the $k_x$-$k_y$ plane evolve from complex shapes to nearly planar sheets at $k_z = \pi/c$ (Fig.~\ref{Fig:FS}(b)). These planar sheets can nest strongly to induce electronic instabilities and symmetry-breaking states such as CDW, as found in our experiments. 

\begin{figure}[t]
\includegraphics[width=\columnwidth]{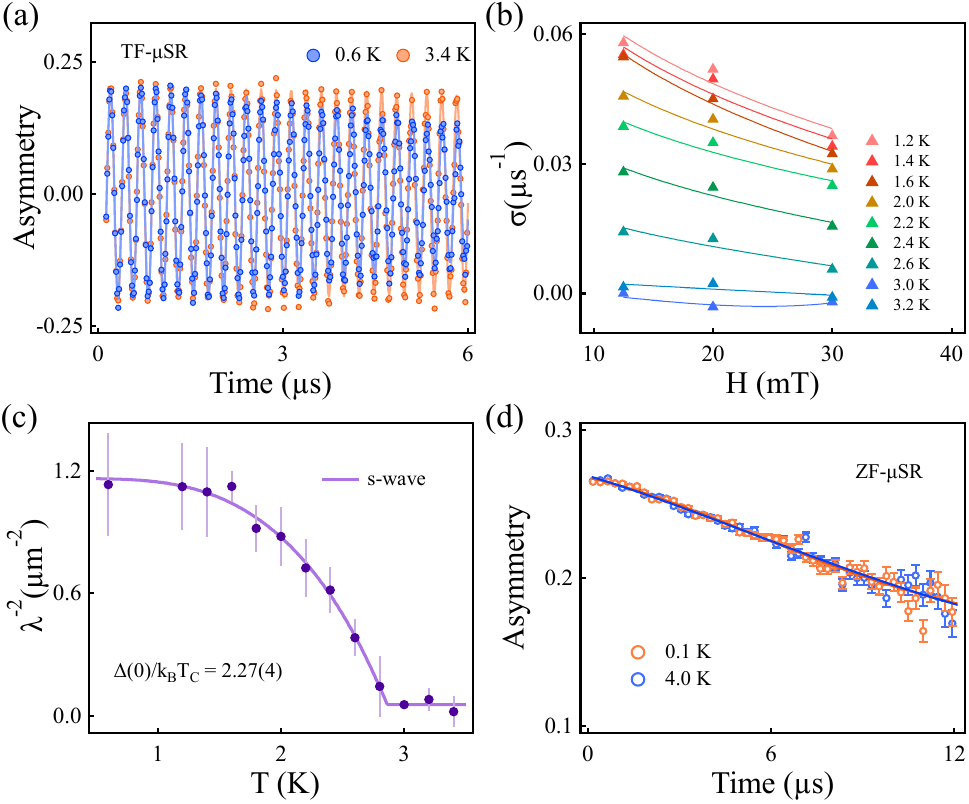}
\caption {\label{Fig:muon} \textbf{Muon spin rotation/relaxation
($\mu$SR) study on \ch{PtPb3Bi}:}
(a) Time-domain transverse field $\mu$SR asymmetry spectra under 12.5 \si{mT} magnetic field above $T_C$ (3.4 \si{K}, orange) and below $T_C$ (0.6 \si{K}, blue). The spectral fits are denoted by solid lines.
(b) The variation of muon spin-depolarization rate with magnetic field for different temperatures, fitted with Brandt's equation.
(c) The obtained variation of fitted parameters $\lambda^{-2}$ with temperature. The violet line represents the best fit for the data.
(d) Zero-field $\mu$SR asymmetry spectra taken at 0.1 \si{K} and 4.0 \si{K}. Solid lines show fitting with a muon relaxation function given by Eq. \ref{eqn:ZF}.}
\end{figure}

\noindent \textit{Transverse field \texorpdfstring{$\mu$}{mu}SR:}
To study the nature of the superconducting energy gap microscopically, transverse field (TF) $\mu$SR measurements were performed on \ch{PtPb3Bi}. Before cooling the sample below the superconducting transition temperature, a constant magnetic field greater than $H_{C1}$(0) of the material (12.5 \si{mT} $\le H \le$ 30 \si{mT}) was applied perpendicular to the initial direction of muon spin. This method was employed to establish a flux line lattice (FLL) in the mixed vortex state of a type-II superconductor, resulting in a regular field distribution. Figure~\ref{Fig:muon}(a) represents the muon asymmetry spectra in the superconducting and normal states. In the superconducting state (0.6 \si{K}), there is a notable decay in the asymmetry spectra amplitude due to an inhomogeneous field distribution of the FLL, compared to the uniform FLL in the normal state (3.4 \si{K}). The TF $\mu$SR asymmetry spectra are nicely fitted by the summation of sinusoidally oscillating functions, damped with a Gaussian relaxation component (see SM for details \cite{SM}).

The Gaussian relaxation rate $\sigma$(T) was deduced from the fitting of the TF asymmetry data, which was analyzed as described in the SM \cite{SM} to determine the values of temperature-dependent $\lambda^{-2}$ as shown in Figure~\ref{Fig:muon}(c). The nature of the superconducting gap was estimated by the best fitting of $\lambda^{-2}$(T) with the dirty limit approximation by London for BCS type superconductors, which can be expressed as:
\begin{equation}
\frac{\lambda^{-2}(T)}{\lambda^{-2}(0)} = \frac{\Delta (T)}{\Delta (0)}\tanh{\left[\frac{\Delta (T)}{2k_BT}\right]}
\label{Eq:swavemuon}
\end{equation}
where $\Delta({T}) = \Delta(0)\tanh[1.82(1.018(\frac{T_c}{T}-1))^{0.51}]$ is the temperature-dependent energy gap under the BCS approximation \cite{Carrington2003MgB2}. The fitting of the data yields an energy gap $\Delta$(0) = 0.59(1) \si{meV}. BCS parameter ${\Delta(0)}/{k_{B}T_{C}}$ of 2.27(4) obtained from TF $\mu$SR measurement is slightly higher than the value of 2.03(1) from specific heat data, although both reveal an isotropic s-wave pairing symmetry of the energy gap. The value of the energy gap indicates the moderate strength of the superconducting pairing mechanism, which is greater than the BCS weak coupling limit. Discrepancy in the energy gap values may be due to the smaller number of $\lambda^{-2}$(T) data points below 1.2 \si{K}.

\noindent \textit{Zero field \texorpdfstring{$\mu$}{mu}SR:}
Absence of time-reversal symmetry (TRS) due to spontaneous magnetic fields in the superconducting state can be very minutely detected using the zero-field mode of muon spin relaxation (ZF $\mu$SR) measurements. ZF asymmetry spectra for \ch{PtPb3Bi} were collected in transverse field configuration, both above (4 \si{K}) and below (0.1 \si{K}) the superconducting transition temperature, as indicated in Figure~\ref{Fig:muon}(d). Random nuclear moments are responsible for muon spin polarization decays when static magnetic or electronic moments are absent. An ordered magnetic structure is also not present, as no oscillatory component is visible in the plot. Depolarization of the muon can theoretically be modeled using the Gaussian Kubo-Toyabe function \cite{hayano1979kubo}:
\begin{equation}
G_{\text{ZF}}(t) = \frac{1}{3}-\frac{2}{3}(\sigma^{2}_{\text{ZF}}t^{2}-1)\mathrm{exp}\left(\frac{-\sigma^{2}_{\text{ZF}}t^{2}}{2}\right),
\label{Eq:kubo}
\end{equation}
where $\sigma_{\text{ZF}}$ denotes the nuclear dipolar field width encountered by the muons. The ZF spectra were fitted by the relaxation function given by Eq. \ref{eqn:ZF} as shown in Figure~\ref{Fig:muon}(d),
\begin{equation}
A(t) = A_{1}G_{\text{ZF}}(t)\mathrm{exp}(-\Lambda t)+A_{\text{BG}},
\label{eqn:ZF}
\end{equation}
where $A_{1}$ and $A_{\text{BG}}$ are the sample asymmetry and the time-independent background asymmetry, respectively. $\mathrm{exp}(-\Lambda t)$ term exhibits any other relaxation channels, such as time-reversal symmetry breaking. 
In the presence of any spontaneous field, ZF spectra show an increase in the relaxation rate in the superconducting region \cite{hillier2022muon,Sajilesh2024hfrhge,singh2018re6ti}. While fitting of ZF spectra for \ch{PtPb3Bi} remains the same in the superconducting and normal regions. Fitting parameters $\Lambda$ (electronic relaxation rate) and $\sigma_{\text{ZF}}$ (nuclear depolarization rate) show no difference within the error limit, as shown by Figure~\textcolor{blue}{S2}(b) in SM \cite{SM}. It suggests that the superconducting ground state of \ch{PtPb3Bi} preserves time-reversal symmetry within the detection range of $\mu$SR.

\noindent \textit{Pairing symmetry:} Finally, we discuss the pairing states that are consistent with our findings. The presence of inversion symmetry leads to doubly degenerate normal-state bands. Despite the significant impact of SOC on the band structure, we can still use a pseudospin basis (Pauli matrices $s_j$), with the same transformation properties as spin, to describe the low-energy subspace close to the Fermi surface at momentum $\vec{k}$. The superconducting order parameter can then be written as $\Delta(\vec{k}) = (\psi_{\vec{k}} s_0 + \vec{d}_{\vec{k}}\cdot \vec{s}) is_y$. Inversion symmetry also implies that the singlet ($\psi$) and triplet ($\vec{d}$) states do not mix. In fact, the aforementioned point group D$_{4h}$ has four inversion-even 1D IR, associated with four different singlet superconductors, and four odd 1D IR, associated with four triplet states. On top, there is one even and one odd 2D IR, each associated with three superconducting states \cite{RevModPhys.63.239}. Out of these 14 candidate orders, only two are consistent with a full gap and thus with our specific heat and $\mu$SR data: the $A_{1g}$ singlet, with $\psi_{\vec{k}} = \delta_1 + \delta_2 (X^2_{\vec{k}} + Y^2_{\vec{k}}) + \delta_3 Z^2_{\vec{k}}$, and the $A_{1u}$ triplet with $\vec{d}_{\vec{k}} = (\delta_1 X_{\vec{k}},\delta_1 Y_{\vec{k}},\delta_2 Z_{\vec{k}})$; here $\delta_j \in \mathbb{R}$ are coefficients and $X_{\vec{k}}$, $Y_{\vec{k}}$, $Z_{\vec{k}}$ real-valued, Brillouin zone-periodic functions transforming, respectively, as $k_x$, $k_y$, $k_z$ under D$_{4h}$. Note that neither of these two states breaks TRS, which is also consistent with our zero-field $\mu$SR data. Given the likely highly disordered nature of our sample (see SM \cite{SM}), a triplet state does not seem to be natural, due to its disorder sensitivity. As such, the $A_{1g}$ singlet is the most likely superconducting state of \ch{PtPb3Bi}, given our observations.

Naturally, the additional presence of the CDW order will reconstruct the band structure and may further reduce the symmetries. However, the analog of the $A_{1g}$ state, transforming trivially under all residual symmetries, still exits and will also be fully gapped and preserve TRS. We further note that the complex unreconstructed Fermi surfaces in Figure~\ref{Fig:FS}(a) make it very likely that CDW order will still leave low-energy spectral weight at the Fermi level, such that pairing is naturally expected at sufficiently low temperatures, in line with our findings.

\noindent \textit{Conclusion:}
In summary, we report a new quasi-1D superconductor \ch{PtPb3Bi}. The compound stabilizes in a nonsymmorphic structure with space group \textit{P4$_2$/mnm}, as confirmed by XRD and EDX analysis. Electrical transport, magnetization, and specific heat measurements establish bulk type-II superconductivity below 3.01(1) \si{K}, observed here for the first time. $\mu$SR measurements probe the superconducting ground state: ZF-$\mu$SR shows preservation of time-reversal symmetry, while TF-$\mu$SR confirms conventional $s$-wave pairing. The superconducting energy gap and electron-phonon coupling constant exceed BCS values, indicating moderately strong coupling. Bulk band structure exhibits strong dispersion along $-Z$--$\Gamma$--$Z$, consistent with the quasi-1D character. Fermi surface comprises multiple pockets and sheets, with planar contours that support strong nesting and a tendency towards electronic instabilities. Electrical resistivity shows a clear CDW transition at 280(1) \si{K}, which could arise due to Fermi surface nesting~\cite{cdweph}. Considering the experimental and theoretical results, the analysis of pairing symmetry suggests that the $A_{1g}$ singlet is the most probable superconducting state for \ch{PtPb3Bi}.

Surface-state calculations along the (010) plane reveal nontrivial states within the bulk gap, consistent with the topology inferred from Wannier charge center evolution. Transport measurements reveal low carrier mobility, which is consistent with the theoretically predicted low intrinsic in-plane mobility. This places \ch{PtPb3Bi} in the dirty limit, with diffusive normal-state transport \cite{disorder,anderson}. Residual resistivity ratio, the Ioffe-Regel parameter, and the ratio between the mean free path and the BCS coherence length further suggest strong electronic correlations, which may endure singlet superconductivity in the highly disordered quasi-1D compound \cite{brahlek2024hidden,abrikosov1983superconductivity}. In such a regime, disorder suppresses long-range CDW order through random pinning and favors short-range correlations, consistent with the observed broad transport anomaly~\cite{abrikosov1983superconductivity,Gruner,imryma}. Table \textcolor{blue}{S1} in SM~\cite{SM} summarizes the extracted parameters, and Fig.~\textcolor{blue}{S4} compares the superconducting critical temperature and upper critical field with other quasi-1D systems. Moderate pairing strength and nontrivial topology support low-energy excitations and unusual magnetotransport properties, and point to \ch{PtPb3Bi} as a candidate for topological superconductivity.

\noindent \textit{Acknowledgments:}
We thank Vivek Kumar Anand, Rajesh Tripathi, and Amitava Bhattacharyya for interesting discussions on $\mu$SR data. S.~S. acknowledges the University Grants Commission (UGC), Government of India, for the Senior Research Fellowship (SRF). R.~P.~S. acknowledges the SERB Government of India for the Core Research Grant No. CRG/2023/000817. The work at TIFR Mumbai is supported by the Department of Atomic Energy, Government of India, under Project Identification Nos. RTI4013 and RTI4015.  The authors thank ISIS, STFC, UK, for providing beam time to carry out the $\mu$SR experiments \cite{isisdoi}. D.~T.~A. would like to thank the Royal Society of London for International Exchange funding between the UK and Japan, Newton Advanced Fellowship funding between UK and China, the CAS for PIFI Fellowship, and EPSRC UK for the funding (Grant No. EP/W00562X/1).

\bibliography{Library}

\clearpage

\onecolumngrid

\begin{center}
    {\Large \textbf{\textrm{Supporting Information for\\ "Discovery of Quasi-One-Dimensional Superconductivity in \ch{PtPb3Bi}"}}}
\end{center}

\vspace{10pt}


\setcounter{figure}{0} 

\renewcommand{\thefigure}{S\arabic{figure}}  

\setcounter{table}{0} 

\renewcommand{\thetable}{S\arabic{table}}  


\twocolumngrid

In the Supplementary Information, we present computational details, comprehensive details about material synthesis and experimental data on electrical transport, magnetization, specific heat, muon spin relaxation/rotation, electronic properties, and Uemura plot for \ch{PtPb3Bi}.\\

\noindent \textit{Computational Details:}
Electronic structure calculations were performed within the framework of first-principles density functional theory (DFT) using the \textit{Vienna Ab initio Simulation Package} (VASP) \cite{kohn1965self,kresse1999paw}. The atomic positions were relaxed until the forces on each atom were smaller than $10^{-4}\,\mathrm{eV}/\text{\AA}$. The generalized gradient approximation (GGA) \cite{perdew1996gga} was employed for the exchange--correlation functional, and the effects of spin orbit coupling (SOC) were included self-consistently. A plane-wave energy cutoff of 350~eV was used. The Brillouin zone was sampled using a $3\times3\times9$ $k$-mesh centered at the $\Gamma$-point for self-consistent calculations. The Fermi surface was generated by constructing a material-specific tight-binding model via the VASP2WANNIER90 interface~\cite{wannier90}. The slab band structure was calculated using the tight-binding method as implemented in the wanniertools package~\cite{wanniertools}.\\

\noindent \textit{Experimental Details:}
The high-purity elemental powders of Pt (4N), Pb (5N), and Bi (5N) were used in the required 1:3:1 ratio. The stoichiometric mixture was ground, palletized, heated at 923 \si{K} for 5 days, and then quenched into cold water at 723 \si{K}. We attempted multiple times to synthesize pure single crystals of \ch{PtPb3Bi}; however, the resulting crystals consistently contained impurities in the form of a Pb–Bi alloy phase \cite{pbbi}. Therefore, we carried out a detailed analysis of the polycrystalline sample. Room-temperature powder X-ray diffraction (XRD) data were collected using a PANalytical X’Pert diffractometer with Cu-K$_\alpha$ radiation. Energy-dispersive X-ray (EDX) analysis was conducted using a scanning electron microscope (SEM). Magnetization measurements were performed using a vibrating sample magnetometer (VSM) in a 7 \si{T} Quantum Design MPMS-3 Superconducting Quantum Interference Device (SQUID) magnetometer. Resistivity measurements were conducted using the four-probe method with a Quantum Design 9 \si{T} Physical Property Measurement System (PPMS). Specific heat measurements using the two-tau model were performed with a dilution refrigerator (DR) attached to a Quantum Design Dynacool PPMS.

Muon spin relaxation/rotation ($\mu$SR) measurements were performed in both zero-field (ZF) and transverse-field (TF) configurations using MUSR at the ISIS neutron and muon source based at Rutherford Appleton Laboratory, United Kingdom, owned and operated by STFC, UKRI. Approximately 3.0 grams of \ch{PtPb3Bi} sample was mounted on the silver holder in powder form using diluted GE varnish and installed in a dilution refrigerator, which can be operated between 35 \si{mK} and 4.2 \si{K}. Zero-field measurements were assured within an error limit of 1 \si{\micro T} using current-carrying coils and an active compensation system to compensate for external stray fields due to the Earth and nearby instruments (technical details \cite{hillier2022muon}).\\

\noindent \textit{Structural Characterization:}
The Rietveld refinement of the powder XRD pattern using \ch{PtPb3Bi} crystallographic information file (CIF) is performed with FullProf Suite software. It confirms the phase purity and the tetragonal crystal structure under the space group \textit{P4$_2$/mnm} (No. 136, point group D$_{4h}$), which is nonsymmorphic and centrosymmetric in nature. The lattice constants obtained from the Rietveld refinement [$a$ = $b$ = 11.4505(1) $\text{\AA}$, $c$ = 4.0839(5) $\text{\AA}$, $V_{cell}$ = 535.463(4) $\text{\AA}^{3}$] agree well with the previously reported values \cite{structure,MATKOVIC1978P35}. EDX analysis further confirms the nominal composition \ch{Pt_{1.03}Pb_{3.08}Bi_{1.00}} and homogeneity of the sample.\\

\noindent \textit{Electrical Resistivity:}
The superconductivity in \ch{PtPb3Bi} is confirmed by the zero-drop in resistivity at the transition temperature $T_{c,offset}$ of 3.08(1) \si{K} (inset of Fig. \textcolor{blue}{1}(b)). The transition temperature width of the zero-drop is 0.6 \si{K}, confirming the bulk superconductivity in the sample, supported by magnetization and specific heat measurements, discussed in the following sections. 

Temperature variation of AC electrical resistivity $\rho(T)$ in the range of 1.9 \si{K} to 400 \si{K} under zero applied magnetic field was performed in both zero-field cooling and warming modes, and a clear bifurcation in the data is depicted in Figure~\textcolor{blue}{1}(b). The difference in cooling and warming data indicates a charge density wave transition at $T_{CDW} = 280(1)$ \si{K}. The normal state resistivity data indicate a metallic behavior below $T_{CDW}$. The value of the residual resistivity ratio [$RRR$=$\rho$(400 K)/$\rho$(5 K)] is 1.19, indicating the presence of scattering and disorder in the sample \cite{Gruner}.\\

\begin{figure*}[t]
\includegraphics[width=0.7\textwidth]{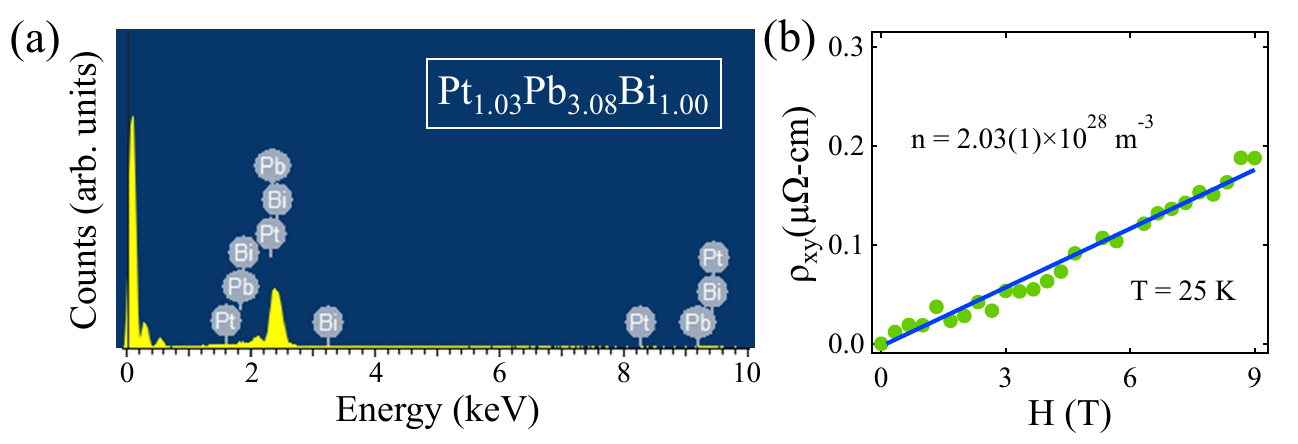}
\caption {\label{Fig:S1}
(a) The EDX spectra for \ch{PtPb3Bi} showing Pt, Pb and Bi elemental peaks. (b) The magnetic field variation of Hall resistivity at 25 \si{K}.}
\end{figure*}

\noindent \textit{Magnetization:}
The magnetization versus temperature measurement confirms the bulk superconductivity in \ch{PtPb3Bi} compound. Magnetization was measured in zero field-cooled warming (ZFCW) and field-cooled cooling (FCC) modes under an applied field of 1 \si{mT}. The superconducting transition temperature obtained from the magnetization measurements is 3.01(1) \si{K}. A strong flux pinning can be observed from the separation of the FCC from the ZFCW curves, providing evidence of type II superconductivity. The superconducting volume fraction of $\sim$100$\%$ was calculated by incorporating the demagnetization factor (N) correction (depending on the shape of the sample and orientation with the applied magnetic field), as discussed in this section. The type II superconducting nature was affirmed by the magnetic field-dependent magnetization loop at 1.8 \si{K}, as illustrated in the inset of Figure~\textcolor{blue}{1}(c), with irreversible magnetic field, $H_{irr}=0.50(4)$ \si{T}.

Moreover, magnetization was measured under a low magnetic field range at different temperatures below $T_C$ (refer to the inset of Fig. \textcolor{blue}{1}(d)). The lower critical field, $H^*_{C1}$ values extracted at each temperature by the deviation of the curves from the Meissner line, give the lower critical fields at different temperatures. The slope of the Meissner line was used to calculate the demagnetization factor (N) using the relation $1/4\pi(1-N)=-M/H$ following ref. \cite{demagcava}. Figure~\textcolor{blue}{1}(d) describes the variation of $H^*_{C1}$ with reduced temperature, which was fitted using the Ginzburg-Landau (GL) equation as,
\begin{equation}
H^*_{C1}(T)=H^*_{C1}(0)[{1-t^{2}}],  \quad  \text{where} \;  t = \frac{T}{T_{C}}.
\label{eqn3:HC1}
\end{equation}
The intersection of the extrapolated fitting curve with the y-axis gives $H^*_{C1}(0) = 1.16(1)$ \si{mT} for \ch{PtPb3Bi}. The demagnetization corrected value of the lower critical field, $H_{C1}(0)=H^*_{C1}(0)/(1-N)$ was found to be 2.09(4) \si{mT}. Similarly, the values of the upper critical field, $H_{C2}$ versus reduced temperature were also well fitted using the GL equation defined by Eq. \ref{eqn4:HC2} (refer to Fig. \textcolor{blue}{1}(e)). The values of $H_{C2}$(T) can be obtained from the temperature variation of both magnetization and resistivity data. Resistivity (and magnetization) curves show a decrease in $T_C$ when the magnetic field is raised, providing the $H_{C2}$ values at various temperatures (Fig. \textcolor{blue}{1}(e), inset).
\begin{equation}
H_{C2}(T) = H_{C2}(0)\left[\frac{1-t^{2}}{1+t^{2}}\right],  \quad  \text{where} \;  t = \frac{T}{T_{C}}.
\label{eqn4:HC2}
\end{equation}
The extrapolation of the $H_{C2}$(T) versus $T/T_C$ data gives an upper critical field $H_{C2}$(0) value of 1.43(1) \si{T} and 1.32(1) \si{T} from the resistivity and magnetization measurements, respectively. $H_{C2}$(0) value shows enhancement from the value of 0.35 \si{T} for the \ch{PtPb4} compound \cite{xu2021superconductivity}.

\begin{figure*}[t]
\includegraphics[width=0.6\textwidth]{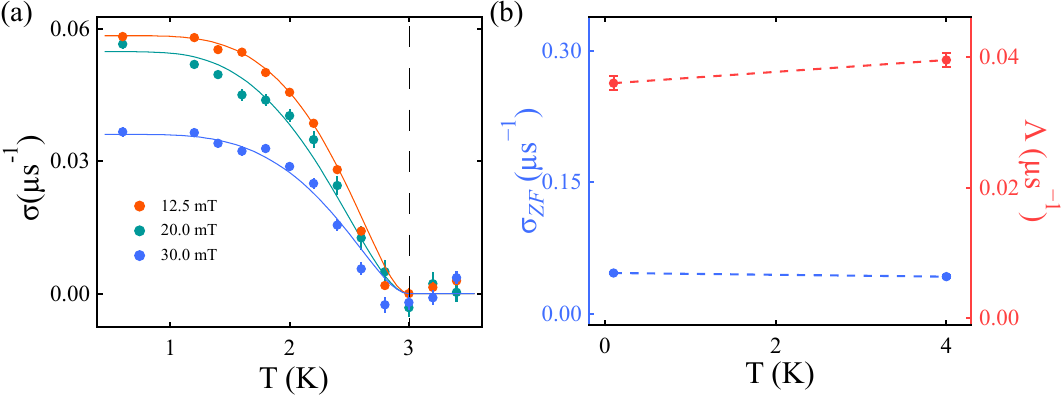}
\caption {\label{Fig:S2} (a) TF-$\mu$SR spin depolarization rate $\sigma$ versus temperature $T$ at applied magnetic fields from 12.5 to 30 \si{mT}. The dashed isothermal line separates the superconducting state and the normal state. The solid lines are a guide to the eye. (b) The unnoticeable changes in electronic relaxation rate $\Lambda$ and nuclear relaxation rate $\sigma_{ZF}$ with temperature within the error range.}
\end{figure*}

Annihilation of superconductivity under the influence of a magnetic field can occur through two distinct mechanisms. (i) spin paramagnetic effect, a Cooper pair breaking caused by the Zeeman effect known as the Pauli limiting field $H_{C2}^{P}$(0) = 1.86 $T_{C}$ \cite{Chandrasekhar1962pauli, Clogston1962pauli}. (ii) orbital limiting effect, caused by the increase in the kinetic energy of the supercurrent exceeding superconducting gap energy, which can be evaluated by the Werthamer-Helfand-Hohenberg (WHH) theory for a type-II superconductor using Eq. \ref{eqn5:WHH} \cite{WHH1966orbital, Helfand1966orbital}.
\begin{equation}
H^{orbital}_{C2}(0) = -\alpha T_{C} \left.{\frac{dH_{C2}(T)}{dT}}\right|_{T=T_{C}}, 
\label{eqn5:WHH}
\end{equation}
where $\alpha$ is a constant called the purity factor, which has different values for dirty and clean limit superconductors (0.69 and 0.73, respectively). The value of $H_{C2}^{P}$(0) for T$_{C}$ of 3.01(1) \si{K} is 5.60(1) \si{T}, and $H^{orbital}_{C2}$(0) was found to be 0.92(1) \si{T} for the dirty limit case. The value of $H_{C2}$(0) is significantly lower than the Pauli paramagnetic limit of 5.60(1) \si{T}, signifying the role of orbital effects in the pair breaking. The impact of orbital effects can be quantified using the Maki parameter, $\alpha_{m}$ = $\sqrt{2}$ $\frac{H_{C2}^{orbital}(0)}{H_{C2}^{P}(0)}$, giving $\alpha_{m}$ of 0.23(2) for the limiting field values of the \ch{PtPb3Bi} compound.

Characteristic length parameters, coherence length, $\xi_{GL}$ and penetration depth $\lambda_{GL}$ can be calculated using the relations $H_{C2}(0) = {\frac{\phi_{0}}{2\pi \xi_{GL}^2(0)}}$ and $H_{C1}(0) = \frac{\phi_{0}}{4\pi\lambda_{GL}^2(0)}\left[ln \frac{\lambda_{GL}(0)}{\xi_{GL}(0)} + 0.12\right]$, where $\phi_{0}$ (magnetic flux quantum) is a constant \cite{tinkham2004introduction}. The values of $\xi_{GL}$ and $\lambda_{GL}$ for the sample were calculated to be 15.8(1) \si{nm} and 534.5(5) \si{nm}, respectively.

The Ginzburg-Landau (GL) theory defined a parameter known as the GL parameter $\kappa_{GL}$ to quantitatively distinguish between type-I and type-II superconductors. The value of $\kappa_{GL}$ = $\frac{\lambda_{GL}(0)}{\xi_{GL}(0)}$ was estimated to be 33.8(4), which is much higher than $1/\sqrt{2}$, confirming that \ch{PtPb3Bi} is a type-II superconductor. The thermodynamic critical field parameter $H_{C}$, can also be assessed using $H_{C1}$(0) and $H_{C2}$(0) \cite{tinkham2004introduction}. Equation $H_{C}^2 ln\kappa_{GL} = {H_{C1}(0) H_{C2}(0)}$ gives $H_{C}$ = 28.0(2) \si{mT}. All superconducting characterization parameters estimated here are outlined in \tableref{tbl: parameters} for the \ch{PtPb3Bi} compound.\\

\noindent \textit{Specific Heat:}
Bulk superconductivity in the \ch{PtPb3Bi} polycrystal is confirmed by using the temperature variation of specific heat $C(T)$ measurements. The significant jump in $C(T)/T$ versus $T^2$ in zero magnetic field manifests the superconducting transition temperature $T_{C}=2.96(4)$ \si{K} of the sample (Fig. \textcolor{blue}{1}(f) inset). The $C(T)/T$ data above $T_C$ is fitted using the Debye-Sommerfeld relation (Eq. \ref{Eq1:Debye}). The first term $\gamma_{n}T$ represents the electronic contribution to the specific heat, and the following terms $\beta_{3}T^{3}$ and $\beta_{5}T^{5}$ denote the phononic and anharmonic contribution, respectively. The inset of Figure~\textcolor{blue}{1}(f) shows the fitting of the low-temperature data with Eq. \ref{Eq1:Debye} yielding the Sommerfeld coefficient $\gamma_{n}$ = 10.31(2) \si{mJmol^{-1}K^{-2}}, the Debye constant $\beta_{3}$ = 4.64(5) \si{mJmol^{-1}K^{-4}}, and $\beta_{5}$ = 0.15(7) \si{mJmol^{-1}K^{-6}}.
\begin{equation}
C = \gamma_{n}T + \beta_{3}T^{3} + \beta_{5}T^{5}.
\label{Eq1:Debye}
\end{equation}
The evaluated value of the density of state at the Fermi level $D_{C}(E_{F})$ is 4.37(2) states eV$^{-1}$f.u.$^{-1}$ using the relation $\gamma_{n}$ = $\left(\frac{\pi^{2} k_{B}^{2}}{3}\right) D_{C}(E_{F})$, where $k_{B}$ $\approx$ 1.38 $\times$ 10$^{-23}$ \si{JK^{-1}}. Debye temperature $\theta_{D}$ can be related to the Debye constant $\beta_{3}$ using the relation $\theta_{D}$ = $\left(\frac{12\pi^{4} R N}{5 \beta_{3}}\right)^{\frac{1}{3}}$, and is calculated to be 127(8) \si{K}. The number of atoms per formula unit $N$ is 5 for \ch{PtPb3Bi} sample, and the universal gas constant R is taken as 8.314 \si{Jmol^{-1}K^{-1}}. The inverted McMillan’s equation \cite{mcmillan1968transition} can be used to calculate the electron-phonon coupling constant $\lambda_{e-ph}$, given as
\begin{equation}
\lambda_{e-ph} = \frac{\mu^{*}\mathrm{ln}(\theta_{D}/1.45T_{C})+1.04}{(1-0.62\mu^{*})\mathrm{ln}(\theta_{D}/1.45T_{C})-1.04};
\label{eqn2:Lambda}
\end{equation}
where $\mu^{*}$ is a repulsive-screened Coulomb pseudopotential parameter that is taken to be 0.13 for intermetallic compounds. The value of $\lambda_{e-ph}$ is 0.71(3) from the measured variables, indicating a moderate strength of electron-phonon coupling, which sustains superconductivity in the presence of a strong disordered quasi-1D material.

\begin{table}[b]
\caption{Superconducting and normal state parameters extracted from magnetization, resistivity, specific heat, and $\mu$SR measurements for \ch{PtPb3Bi}.}
\label{tbl: parameters}
\setlength{\tabcolsep}{14pt}
\renewcommand{\arraystretch}{1.3} 
\begin{center}
\begin{tabular}[b]{lcc}\hline \hline
Parameters                                  & unit                  & \ch{PtPb3Bi}  \\
\hline
$T_{C}^{mag}$                               & \si{K}                & 3.01(1)       \\
$H_{C1}(0)$                                 & \si{mT}               & 2.09(4)       \\ 
$H_{C2}^{res}$(0)                           & \si{T}                & 1.43(1)       \\
$T_{CDW}$                                   & \si{K}                & 280(1)        \\
$\xi_{GL}$                                  & \si{nm}               & 15.8(1)       \\
$\lambda_{GL}^{mag}$                        & \si{nm}               & 534.5(5)      \\
$\kappa_{GL}$                               & -                     & 33.8(4)       \\
$\alpha_{m}$                                & -                     & 0.23(2)       \\
$\gamma_{n}$                                & \si{mJmol^{-1}K^{-2}} & 10.31(2)      \\
$\theta_{D}^{sp}$                           & \si{K}                & 128(1)        \\
$\frac{\Delta(0)}{k_{B}T_{C}}$ (sp. heat)   &  -                    & 2.03(1)       \\
$\frac{\Delta(0)}{k_{B}T_{C}}$ ($\mu$SR)    &  -                    & 2.27(4)       \\
$\lambda_{e-ph}$                            & -                     & 0.71(3)       \\
$D_{C}(E_{F})$                              & states/(eV f.u.)      & 4.37(2)       \\
$m^{*}/m_{e}$                               & -                     & 2.9(1)        \\
$n$                                         & 10$^{28}$ \si{m^{-3}}            & 2.03(1)       \\
v$_{F}$                                     & 10$^{5}$ m/s          & 3.4(1)        \\
$\xi_{0}$                                   & \si{nm}               & 152.8(8)      \\
$l_e$                                       & \si{nm}               & 0.27(5)       \\
$\mu_m$                                       & \si{cm^2V^{-1}s^{-1}}               & 0.35(1)       \\
$T_C/T_F$                                       & -                & 0.00028     \\
\hline \hline
\end{tabular}
\par\medskip\footnotesize
\end{center}
\end{table}

The variation of the normalized electronic specific heat with the reduced temperature was measured to analyze the nature of the superconducting gap symmetry of the sample. Electronic-specific heat $C_{el}$ can be estimated by subtracting phononic and anharmonic contributions from the total specific heat $C$. The jump in normalized specific heat $\frac{\Delta C_{el}}{\gamma_{n}T_C}$ is 1.64(2), which is higher than the weak coupling limit from the BCS theory (i.e., 1.43). Further, Figure~\textcolor{blue}{1}(f) shows the fitting of $C_{el}/\gamma_{n}T$ versus reduced temperature $T/T_C$ with the isotropic single-gap BCS model given by Eq. \ref{Eq:swave} \cite{padamsee1973quasiparticle}. Models other than the s-wave model fail to properly describe the low-temperature specific heat behavior of \ch{PtPb3Bi}. The equation relates the entropy $S_{el}$ to the temperature-dependent BCS energy gap $\Delta(t) = tanh[1.82\{1.018(\frac{1}{t}-1)\}^{0.51}]$. The entropy S$_{el}$ is related to C$_{el}$ by C$_{el}$ = $t\frac{dS_{el}}{dt}$, where t = $\frac{T}{T_{C}}$ is the reduced temperature.
\begin{equation}
\begin{split}
\frac{S_{el}}{\gamma_{n} T_{C}}= -\frac{6}{\pi^{2}} \left(\frac{\Delta(0)}{k_{B} T_{C}}\right) \int_{0}^{\infty}[ &f_{y}ln(f_{y})\\ 
&+(1-f_{y})ln(1-f_{y})] dy,
\end{split}
\label{Eq:swave}
\end{equation}
where the Fermi function $f_{y}(\xi)$ = $[1+e^{\frac{E(\xi)}{k_{B}T}}]^{-1}$ is integrated with respect to the Fermi energy $y$ = $\xi/\Delta(0)$. The normal electron energy with respect to the Fermi energy is given as $E(\xi)$=$\sqrt{\xi^{2}+\Delta^{2}(t)}$. The best fit of the normalized specific heat data results in a superconducting energy gap $\frac{\Delta(0)}{k_{B}T_{C}}=2.03(1)$ that is higher than the weakly coupled BCS gap value of 1.76. The values of ${\Delta C_{el}}/{\gamma_{n}T_C}$ and ${\Delta(0)}/{k_{B}T_{C}}$ exceeding the BCS values indicate moderately coupled superconductivity in \ch{PtPb3Bi}.\\

\begin{figure}[t]
\includegraphics[width=\columnwidth]{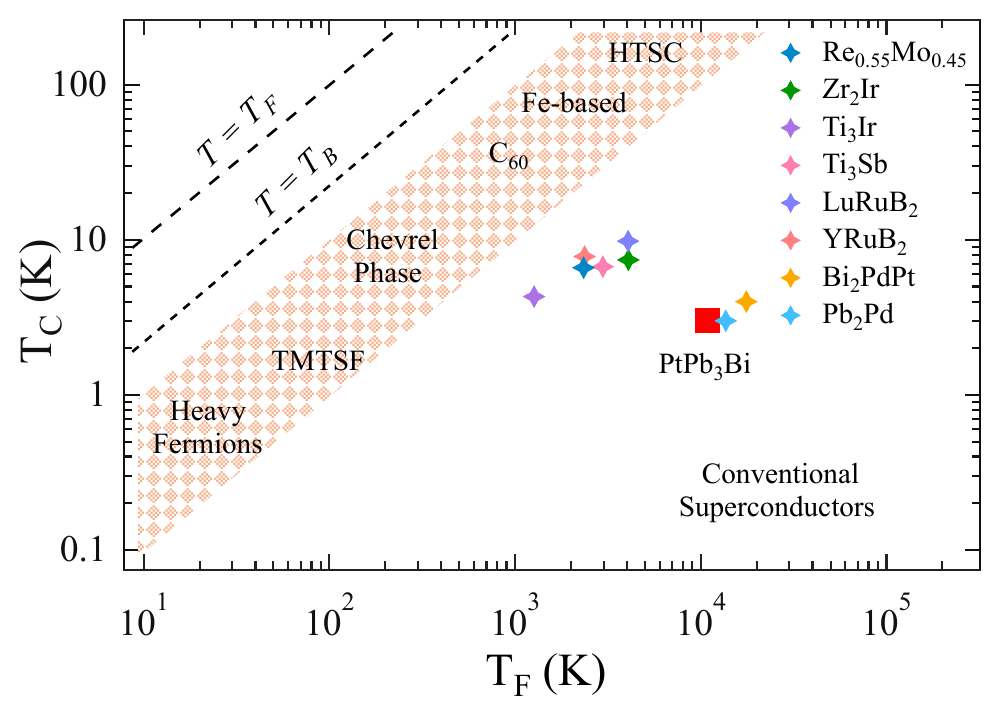}
\caption {\label{Fig:S3} The Uemura plot, superconducting critical temperature versus Fermi temperature, distinguishes the unconventional superconductors from the conventional BCS superconductors based on the ratio between $T_C$ and $T_F$. The red square marker denotes \ch{PtPb3Bi}, which is away from the unconventional region.}
\end{figure}

\noindent \textit{Transverse-Field \texorpdfstring{$\mu$}{mu}SR:}
The TF $\mu$SR asymmetry spectra given in the main text were nicely fitted by the summation of sinusoidally oscillating functions, where each function is damped with a Gaussian relaxation component \cite{maisuradze2009comparison,weber1993flux}:
\begin{equation}
G(t) = \sum_{i=1}^N A_{i}\exp\left(-\frac{1}{2}\sigma_i^2t^2\right)\cos(\gamma_\mu B_it+\phi),
\label{Eq:TF}
\end{equation}
where $\phi$, $A_{i}$, and $\sigma_{i}$ are the initial phase, initial asymmetry, and the Gaussian relaxation rate, respectively. The value of the muon gyromagnetic ratio $\gamma_{\mu}/2\pi$ is 135.5 \si{MHz/T}, and $B_i$ is the $i$th component of the Gaussian field distribution. The asymmetry spectra for the sample were best described by N = 2, representing two Gaussian components with fixed $\sigma_{2} = 0$ to explain the non-depolarizing muon background stopped in the silver sample holder. Thus, $A_{2}$ (and $B_{2}$) account for the background asymmetry (and magnetic field), respectively.

The background nuclear dipolar contribution to the relaxation rate $\sigma_N$ (considered to be temperature-independent for the temperatures discussed) can be removed from the total Gaussian relaxation rate $\sigma_{total}$ to obtain the flux line lattice component of the Gaussian relaxation rate $\sigma$ using equation $\sigma = \sqrt{\sigma_{total}^{2} - {\sigma_N}^{2}}$. The deduced values of the superconducting contribution of the Gaussian relaxation rate $\sigma$(T) for different applied fields (12.5, 20, and 30 \si{mT}) are shown with the error bars in Figure~\ref{Fig:S2}(a). The field-dependent $\sigma$(H) at different temperatures is depicted in Figure~\textcolor{blue}{4}(b), which was extracted from the isothermal values of $\sigma$(T) from Figure~\ref{Fig:S2}(a). Brandt's equation \cite{Brandt} described the field dependence of the penetration depth $\lambda$ in an isotropic type II superconductor with $\kappa_{GL}>5$ by Eq. \ref{Eq:Brandt}, where $h = H/H_{C2}(T)$ is the reduced magnetic field.
\begin{equation}
\sigma(\mu s^{-1}) = 4.854 \times 10^{4}(1-h)[1+1.21(1-\sqrt{h})^{3}]\lambda^{-2}.
\label{Eq:Brandt}
\end{equation}
This relation was used to fit the isothermal datasets in Figure~\textcolor{blue}{4}(b) and extract the corresponding values of temperature-dependent $\lambda^{-2}$, which are presented in the main text, Figure~\textcolor{blue}{4}(c).\\

\begin{figure}[b]
\includegraphics[width=\columnwidth]{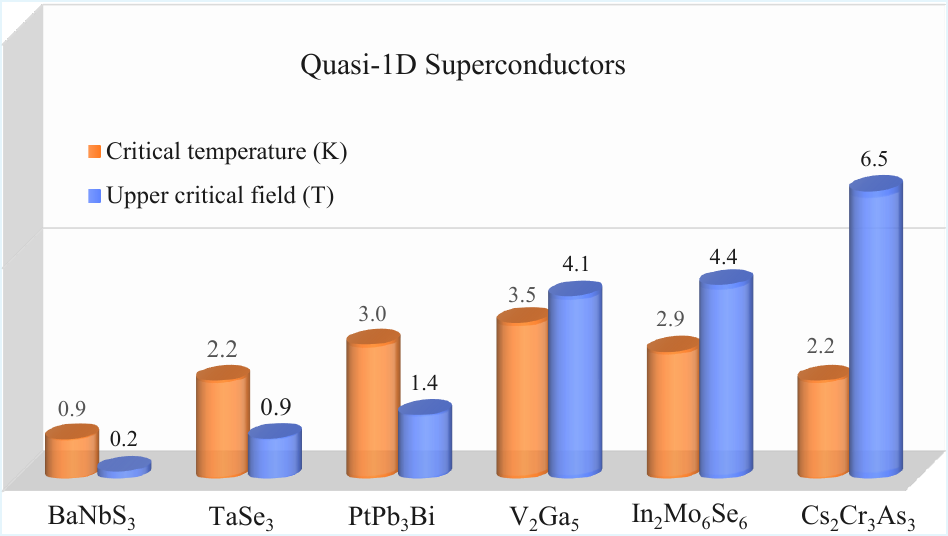}
\caption {\label{Fig:chart} Superconducting transition temperature and upper critical field of \ch{PtPb3Bi}, compared to some other quasi-1D superconductors \cite{Cs2Cr3As3,M2Mo6Se6,BaNbS3,TaSe3,V2Ga5}.}
\end{figure}

\noindent \textit{Electronic Parameters and Uemura Plot:}
The normal state behavior of the charge carriers and the dependent electronic parameters were determined by performing Hall measurements on \ch{PtPb3Bi}. The data for Hall resistivity $\rho_{xy}$ versus applied magnetic field is linearly fitted at 25 \si{K}, as shown in Figure~\ref{Fig:S1}(b). The slope of fitting provides the value of the Hall coefficient $R_H=3.07(9)\times10^{-8}$ \si{\ohm cm T^{-1}}. The sign of $R_H$ indicates holes as charge carriers with a concentration of $n=1/eR_{H}=2.03(1)\times10^{28}$ \si{m^{-3}}.
A superconductor can be designated to be in a clean or dirty limit based on the ratio of BCS coherence length $\xi_{0}=0.18\hbar v_{F}/k_{B}T_C$, and the mean free path $l_{e}= 3\pi^{2}{\hbar}^{3}/e^{2}\rho_{0}m^{*2}v_{F}^{2}$ \cite{tinkham2004introduction}. To calculate these parameters, the value of Fermi wave vector $k_F=(3\pi^2n)^{1/3}=8.4(3)$ \si{nm^{-1}}, effective mass $m^{*}=\gamma_n(\hbar k_F)^2/\pi^2nk_B^2=2.9(1)m_{e}$ and Fermi velocity $v_{F}=\hbar(3\pi^2n)^{1/3}/m^*=3.4(1)\times10^{5}$ \si{m s^{-1}} was evaluated using the value of $\gamma_n$ and $n$ from the specific heat and Hall measurement, respectively. The large value of the ratio between the parameters $\xi_{0}=152.8(8)$ \si{nm} and $l_{e}=0.27(5)$ \si{nm} indicates dirty limit superconductivity likely due to imperfections and disorders. The value of the Ioffe–Regel parameter, $k_Fl_e=2.32(2)$, which is slightly above the Ioffe–Regel limit ($\sim1$), suggests a strongly disordered, diffusive metallic regime \cite{ioffe1960non,disorder}. Carrier mobility $\mu_m=1/ne\rho$ can be calculated from the $\rho$ and $n$ from the resistivity and Hall measurements, respectively. The obtained value of $\mu_m=0.35(1)$ \si{cm^2V^{-1}s^{-1}} is very low, placing the system in the dirty limit with diffusive normal state transport \cite{disorder,anderson}.

Uemura et al. formulated a way to distinguish unconventional superconductors from conventional ones \cite{Uemura}. If the ratio of $T_C$ and $T_F$ is between 0.1 and 0.01, the superconductor is in the unconventional region, which is shown in Figure~\ref{Fig:S3} with a highlighted band comprising notable unconventional superconductor families. The Fermi temperature $T_F=(3\hbar^3\pi^2n)^{2/3}/2m^*k_B$ was calculated to be 10799(189) \si{K} \cite{hillier1997classification}. The ratio of $T_C$ and $T_F$ for \ch{PtPb3Bi} was found to be 0.00028, which lies out of the unconventional band and is shown with a red square marker. Some nonsymmorphic centrosymmetric compounds have been displayed for comparison purposes.

\end{document}